\documentclass[showpacs,twocolumn,preprintnumbers,amsmath,amssymb,prl]{revtex4}
\usepackage{amsmath}
\usepackage{dcolumn}
\usepackage{bm}
\usepackage{graphicx}
\usepackage{subfigure}

\newcommand{\N}{\mathbb{N}}

\newcommand{\One}{\openone}
\newcommand{\ke}[1]{| #1 \rangle}
\newcommand{\br}[1]{\langle #1 |}
\newcommand{\bk}[2]{\langle #1 | #2 \rangle}
\newcommand{\pr}[2]{|#1\rangle \langle #2|}
\newcommand{\moy}[3]{\langle #1 | #2 | #3 \rangle}

\begin{document}
%\preprint{MPIPKS/11-2001}
%\draft

\title{Dynamical Response of Nanomechanical Oscillators in Immiscible Viscous Fluid for {\it in vitro} Biomolecular Recognition}

\author{Jerome Dorignac${}^{1,2}$, Agnieszka Kalinowski${}^3$, Shyamsunder Erramilli${}^{1,3}$,
Pritiraj Mohanty${}^1$}

\affiliation{
${}^1$ Department of Physics, Boston University, 590 Commonwealth Avenue,
Boston, MA 02215\\
${}^2$ College of Engineering, Boston University, 44 Cummington Street, Boston,
  MA 02215\\
${}^3$ Department of Biomedical Engineering, Boston University,
48 Cummington Street, Boston, MA 02215}

\date{\today}

\begin{abstract}

Dynamical response of nanomechanical cantilever structures immersed in a viscous fluid is important to {\it in vitro} single-molecule force spectroscopy, biomolecular recognition of disease-specific proteins, and the detection of microscopic dynamics of proteins. Here we study the stochastic response of biofunctionalized nanomechanical cantilevers beam in a viscous fluid. Using the fluctuation-dissipation theorem we derive an exact expression for the spectral density of the displacement and a linear approximation for the resonance frequency shift. We find that in a viscous solution the frequency shift of the nanoscale cantilever is determined by surface stress generated by biomolecular interaction with negligible contributions from mass loading.

\end{abstract}

%\keywords{Cantilever beams; Viscous fluids; Fluctuation-dissipation}
\maketitle

%\vskip2pc
%\narrowtext

%\section{Introduction} \label{Intro}
From single molecule force spectroscopy\cite{viani} to biomolecular recognition of disease-specific proteins such as cancer antigens \cite{Majumdar01}, micron-sized cantilevers have proved to be fundamental to the ultrasensitive detection of small forces. Usually, forces are detected by measuring the deflection of the cantilever. In the dynamic case, the shift in the resonance frequency of the cantilever is used to infer the magnitude of the force. Micromachining techniques now enable commercial production of such cantilevers with dimensions on the order of 100 $\mu$m as well as their routine use in force spectroscopy. 

Decreasing the cantilever dimensions to sub-micron or nanometer scales increases the resonance frequency to the megahertz-gigahertz range. The resultant increase in the dynamic range and the measurement speed can provide a better tool for probing single molecules.  This could be also used for more sensitive bioimaging techniques and monitoring
real-time binding kinetics of ligand-protein binding as well as the energy landscape of the molecular bonds at their true characteristic time scales. For biomolecular recognition in a viscous fluid, force sensitivity can be increased by decreasing the effective viscous damping. Nanoscale cantilevers are hence expected to have dramatically enhanced force sensitivity as smaller cantilevers have lower viscous damping.

In spite of the obvious importance of nanomechanical cantilevers for ultrasensitive {\it in vitro} force detection, there is no widely-accepted description that relates resonance frequency change to concentration or mass-loading, over the entire range of viscosity, relevant to biomolecular recognition in viscous fluids. In this Letter, we derive an exact expression that takes into account the hydrodynamics of the beam in a continuum approximation with axial loading. We find that, in air, the first mode of the nanoscale cantilever is more sensitive to surface stress than the higher order modes. The frequency shift is primarily determined by surface stress in the first mode, where the mass-loading effects become relevant for higher order modes. More importantly, we find that, in a viscous solution such as water, frequency shift is dominated by surface stress and not mass loading, as generally expected.

%\section{Model} \label{Sec2}

{\it Model:} In standard fluid dynamics, an inviscid model is typically a valid assumption.
However, nanoscale cantilevers have Reynolds numbers less than one, therefore viscous effects become dominant \cite{Purcell}. Proper inclusion of viscous effects is characterized as being in the Stokes-Purcell Regime of fluid flow.

Neglecting rotatory inertia, shear deformation and internal damping, 
the equation of motion for the deflection $y(x,t)$ of a beam
with length $L$, width $b$ and thickness $d$, immersed in a fluid 
at temperature $T$ and loaded by a constant axial force $S$, 
is given by \cite{Humar} 
\begin{equation} \label{Saderseq}
EI\frac{\partial^4 y}{\partial x^4}-S\frac{\partial^2 y}{\partial x^2}+
\mu(x) \frac{\partial^2 y}{\partial t^2} = f_{\rm h}(x,t) + 
f_{\rm th}(x,t).
\end{equation} 
$E$ and $I$ are the Young's modulus and moment 
of inertia of the (coated) beam, respectively. $f_{\rm h}(x,t)$ is
the hydrodynamic loading due to the motion of the fluid around the beam
and $f_{\rm th}(x,t)$ is a Langevin-type force per unit length 
responsible for the thermalization of the beam. The linear mass 
(mass per unit length) of the system $\mu(x)$ consists of the
linear mass of the beam $\mu_b$, and the linear mass
of the trapped biomolecules $\mu_l (x)$. The axial load $S$ introduced in 
\eqref{Saderseq} describes the mutual interaction of biomolecules adsorbed on the beam 
\cite{Thundat95}. 
The boundary conditions for Eq. \eqref{Saderseq} are given by
$y(0,t)=y'(0,t)=0$, $y''(L,t)=0$ and $EIy'''(L,t)=Sy'(L,t)$, 
where primes denote spatial derivatives \cite{Humar}.  

At higher concentrations, the biomolecules form a uniform layer 
with mass $\mu_l$ and thickness $h$ so that
\begin{equation} \label{linmass2}
\mu(x) = \mu_b + \mu_l = constant.
\end{equation}
The mutual interaction of biomolecules within
the layer is modeled by taking into account the stress $\sigma$ 
they generate on the coated surface of the beam. 
As shown in ref.\cite{Majumdar01}, this stress enables the bending of the silicon-nitride microcantilevers with length to thickness ratio $L/d$,
ranging from $10^2$ to $10^3$. The resulting static deflection,
on the order of a few tenths of microns, is related to
surface stress by Stoney's formula \cite{Stoney09}. 
However, for the silicon nanomechanical cantilevers 
under investigation here ($L/d\sim 50$), 
Stoney's formula typically yields angstr\"om-level bendings ($10^{-5}$L). 
So these nanocantilevers remain almost straight under the influence 
of surface stress. Nevertheless, as shown in \cite{Thundat95},
this stress induces an effective axial load, $S=\sigma L$, that must be
included in Eq. \eqref{Saderseq}. In vacuum, such a model has been 
studied in ref.\cite{Lu01}.

In addition, biomolecular interaction on the surface results in an 
effective Young's modulus of the layer $E_l$ [see ref.\cite{Gere}]:
\begin{equation} \label{EI}
EI = E_bI_b+E_lI_l \simeq E_b b d^3/12 + E_l h b d^2/4.
\end{equation} 
$E_bI_b$ and $E_lI_l$ are the respective 
bending rigidities of the beam and the layer (the last equality holds
when $h \ll d$). 
%--------
%  $E_l$ can be experimentally determined from 
%the deflection of the coated nanobeam under external static load. As it is mass-independent, 
% this deflection yields immediately $E_l$. 
% -----------
In the opposite limit, biomolecules with mass 
$m$ sparsely scattered over the beam at locations $x_i$, result in
\begin{equation} \label{linmass1}
\mu(x) = \mu_b + \mu_l (x) = 
\mu_b + m \sum_i \delta(x-x_i), \ \ \ x_i \in [0,L].
\end{equation} 
If the average spacing is large compared to their size,
their mutual interaction is negligible and $S=0$. 
Considering that their presence 
does not substantially affect the moment $I_b$ of the beam, the bending
rigidity of the whole system is the same as for an unloaded beam,
$EI = E_bI_b$.

%\section{Equations of motion} \label{Sec3}
%----------------------------------------------------------------------------

{\it Equations of Motion:} To solve equation \eqref{Saderseq}, 
we expand the deflection 
$y(x,t)$ and the force densities $f_{\rm h}(x,t)$ and $f_{\rm th}(x,t)$ in terms of  
the modes of the {\em bare} beam, defined as the
beam without the added mass ($\mu_l(x)=0$) though it includes 
the tension $S=\sigma L$:  
\begin{equation} \label{expandymodes}
y(x,t)=\sum_{n=1}^{\infty} y_n(t)\phi_n(\xi),\ \ \xi=x/L.
\end{equation}
Similar expressions hold for the force densities. 
The eigenmodes $\phi_n(\xi)$ satisfy the following conditions:
\begin{eqnarray} \label{modes}
&& \!\!\!\!\!\! \phi^{(iv)}_n(\xi)-\alpha \phi_n''(\xi)=\beta_n^4 \phi_n(\xi)\ ;\ \alpha=\sigma L^3/EI; \nonumber \\
&& \!\!\!\!\!\! \phi_n(0)=\phi_n'(0)=\phi_n''(1)=0,\ 
\phi_n'''(1)=\alpha \phi_n'(1).
\end{eqnarray} 
The self-adjointness of \eqref{modes} makes the modes orthonormal:
%\begin{equation} \label{norm}
$\int_0^1\!\!\!\phi_n(\xi)\phi_l(\xi)\, d\xi = \delta_{n,l}$.
%\end{equation} 
From Eq. \eqref{modes},
the eigenvalues $\beta_n$ are the successive positive roots of 
\begin{equation} \label{betan}
1+(1+\varepsilon_n^2)\cosh \lambda_n^+ \cos \lambda_n^- + 
\varepsilon_n \sinh \lambda_n^+ \sin \lambda_n^- = 0 
\end{equation}
where $\varepsilon_n=\alpha/(2\beta_n^2)$ and 
$\lambda_n^{\pm}=
\beta_n[\sqrt{1+\varepsilon_n^2}\pm \varepsilon_n]^{\frac{1}{2}}$.
Note that, for $\varepsilon_n=\alpha/(2\beta_n^2)\ll 1$, 
Eq. \eqref{betan} reduces to the usual clamped-free equation, 
$1+\cos \beta_n \cosh \beta_n=0$. As $\beta_n \propto n$ for large $n$,
{\em eigenvalues for which 
$n \gg \sqrt{\alpha}$ are essentially independent of the surface stress.} 

\noindent
Let $\mu(x)=\mu_b+\mu_l(x)$.
Using \eqref{expandymodes}, Eq. \eqref{Saderseq} reduces to
\begin{equation} \label{Sadersmodes}
M \ddot{y}_n(t) + k_n y_n(t) + \sum_{j=1}^{\infty}\Phi_{nj}\ddot{y}_j(t)
 = F_{n,{\rm h}}(t)+F_{n,{\rm th}}(t). 
\end{equation}
$M=\mu_b L$ is the 
mass of the beam and $F_{n,{\rm h(th)}}(t) = L f_{n,{\rm h(th)}}(t)$. 
The effective stiffness of mode $n$ is 
\begin{equation} \label{kn}
k_n=EI\beta_n^4/L^3.
\end{equation} 
The real and symmetric matrix $\bm{\Phi}$ has components
\begin{equation} \label{Massmatrix} 
\Phi_{nj}=L\int_0^1 \mu_l(\xi'L)\phi_n(\xi')\phi_j(\xi')d\xi'. 
\end{equation}
If $\mu_l(x)=\mu_l$, then $\bm{\Phi}=M_l \bm{\One}$, where $\bm{\One}$
is the identity matrix and $M_l=L\mu_l$ is the layer mass. Eqs. 
\eqref{Sadersmodes} decouple and the mass
of modes $y_n$ becomes the total mass of the system, 
$M+M_l$. But non-uniform mass distributions as in 
Eq. \eqref{linmass1} couple the bare modes of the beam. 

Taking the Fourier transform of \eqref{Sadersmodes} 
and using the expression
for the hydrodynamic force \cite{Rosenhead},
\begin{equation} \label{hydroforce}
\hat{F}_{n,{\rm h}}(\omega) = M_{\!f} \omega^2 \Gamma(\omega) \hat{y}_n(\omega),
\end{equation}
where $M_{\!f}=\frac{\pi}{4}L\rho_f b^2$ is the mass 
of the fluid loading the beam,
and $\Gamma(\omega)=\Gamma_r(\omega)+i\Gamma_i(\omega)$ is a complex 
``hydrodynamic function'' discussed in detail in \cite{theory_Sader},  
we obtain
\begin{equation} \label{Sadersomega}
\bm{\Lambda(\omega)} \ke{\hat{y}(\omega)} = \ke{\hat{F}_{{\rm th}}(\omega)}.
\end{equation}
Kets $\ke{v}$ are column vectors with components 
$v_i,\ i \in \N$. The {\em nonhermitian} matrix $\bm{\Lambda(\omega)}$ 
is given by
\begin{equation} \label{Lambdamat}
\bm{\Lambda(\omega)} = \bm{\Lambda_0(\omega)}-\omega^2\tilde{\bm{\Phi}}
\end{equation}
where
\begin{eqnarray} \label{Lambda0mat}
&& \Lambda_0(\omega)_{nj} = 
[k_n-\omega^2(M_{\!f}\Gamma(\omega)+M_n)]\delta_{nj}
\nonumber \\
&& \tilde{\Phi}_{nj} = \Phi_{nj}(1-\delta_{nj})\ ;\  
M_n = M+\Phi_{nn}. \nonumber 
\end{eqnarray}

%\section{Spectral densities} \label{sec4}
%-----------------------------------------------------------------

{\it Spectral Densities:}
As the {\em dissipative} (complex) part of the hydrodynamic function is 
frequency-dependent, we apply the generalized fluctuation-dissipation
theorem \cite{Smith} to derive the power spectrum matrix of the 
stochastic forces $F_{n,{\rm th}}$, 
$\bm{S_{\hat{F}}(\omega)}=\overline{\pr{\hat{F}_{\rm th}(\omega)}
{\hat{F}_{\rm th}(\omega)}}^s$ 
 (the over-line denotes thermal
averaging, the superscript $s$ refers to the spectral density 
and $\br{\hat{F}_{\rm th}}$
is the hermitian conjugate of $\ke{\hat{F}_{\rm th}}$):
\begin{equation} \label{specdensFF}
\bm{S_{\hat{F}}(\omega)}=\frac{k T}{i\omega}
\left(\bm{\Lambda^{\dagger}(\omega)}-\bm{\Lambda(\omega)}\right) = 
2k T M_{\!f} \omega \Gamma_i(\omega)\bm{\One}
\end{equation} 
where $k$ is the Boltzmann constant and $T$ the temperature.
In components, this yields
\begin{equation} \label{specdensF}
\overline{\hat{F}_{n,{\rm th}}(\omega)
\hat{F}^*_{p,{\rm th}}(\omega')} = 
2k T M_{\!f} \omega \Gamma_i(\omega)\delta_{np}\delta(\omega-\omega').
\end{equation} 
Notice that this expression does not depend on $\bm{\Phi}$. 
It is the same as for a bare beam. As seen above, the stochastic 
forces acting on distinct modes are uncorrelated. Nevertheless, their power 
spectrum cannot be assumed to be constant, contrary to the assumption 
in ref. \cite{theory_Sader},
as the dissipative part of the hydrodynamic function is frequency-dependent.
Eq. \eqref{specdensF} is the generalization of the expression 
derived by Paul and Cross for a single cantilever mode \cite{Paul}. 
Now, inverting \eqref{Sadersomega}, we obtain 
\begin{equation} \label{Chimat}
\ke{\hat{y}(\omega)} = \bm{\chi(\omega)} \ke{\hat{F}_{\rm th}(\omega)},
\ \  \bm{\chi(\omega)} = \bm{\Lambda^{-1}(\omega)}
\end{equation} 
and $\pr{\hat{y}(\omega)}{\hat{y}(\omega)}=\bm{\chi(\omega)}
\pr{\hat{F}_{\rm th}(\omega)}{\hat{F}_{\rm th}(\omega)}
\bm{\chi^{\dagger}(\omega)}$. Spectral averaging the latter 
and using \eqref{specdensFF},
we find the power spectrum matrix of the deflection modes
\begin{equation} \label{Syy}
\bm{S_{\hat{y}}(\omega)}=
2k T M_{\!f} \omega \Gamma_i(\omega)\bm{\chi(\omega)}
\bm{\chi^{\dagger}(\omega)}.
\end{equation} 
Introducing $\ke{\phi_{\xi}}$ 
%and $\ke{\phi'_{\xi}}$ 
with components $\phi_n(\xi)$, 
%and $d\phi_n(\xi)/d\xi$, respectively, 
the Fourier transform of the deflection \eqref{expandymodes} reads 
$\hat{y}(x,\omega)=\bk{\phi_{\xi}}{\hat{y}(\omega)}$ 
%\begin{equation} \label{yanddyofwx}
%\hat{y}(x,\omega) = \sum_{n=1}^{\infty}\phi_n(\xi)\hat{y}_n(\omega) = 
%\bk{\phi_{\xi}}{\hat{y}(\omega)}
%\end{equation} 
and using \eqref{Syy}, we find its spectral density to be
\begin{equation} \label{specdensy}
\overline{\left|\hat{y}(x,\omega)\right|^2}^s =2k T M_{\!f} 
\omega \Gamma_i(\omega) \moy{\phi_{\xi}}{\bm{\chi(\omega)}
\bm{\chi^{\dagger}(\omega)}}{\phi_{\xi}}.
\end{equation}
%Similarly, the spectral density of the slope is given by
%\begin{equation} \label{specdensdy}
%\!\overline{\left|\frac{d\hat{y}}{dx}(x,\omega)\right|^2}^s = 
%\frac{2k T M_{\!f}}{L^2} \omega \Gamma_i(\omega) 
%\moy{\phi'_{\xi}}{\bm{\chi(\omega)}
%\bm{\chi^{\dagger}(\omega)}}{\phi'_{\xi}}.
%\end{equation}
%
The total mass of the particles trapped on the beam $M_l$ is small 
compared to the mass of the beam. This justifies treating 
$\omega^2\tilde{\bm{\Phi}}$ perturbatively provided its elements
stay small compared to the diagonal elements of $\Lambda_0(\omega)$.
This is indeed the case provided $M_{\!f} \Gamma_i(\omega) \gg 4 M_l$ \cite{footnote}.
By inverting \eqref{Lambdamat}, 
we obtain in first order in $\tilde{\bm{\Phi}}$
\begin{equation} \label{chipert}
\bm{\chi}\bm{\chi^{\dagger}} = \bm{\chi_0}\bm{\chi_0^{\dagger}} + 
\omega^2 \bm{\chi_0}\left(\tilde{\bm{\Phi}}\bm{\chi_0}+
\bm{\chi_0^{\dagger}}\tilde{\bm{\Phi}}\right)
\bm{\chi_0^{\dagger}} + {\cal O}(\tilde{\bm{\Phi}}^2)
\end{equation}  
where $\bm{\chi_0}=\bm{\Lambda^{-1}_0}$. 
Reinstating in \eqref{specdensy}, we finally get
\begin{multline} \label{specdensy2}
\overline{\left|\hat{y}(x,\omega)\right|^2}^s =2k T M_{\!f}
\omega \Gamma_i(\omega) \left[\, \sum_{n=1}^{\infty}\frac{
\phi^2_n(\xi)}{M_n^2 |A_n|^2}\ +  
\right. \\ \left. 
+ 2\omega^2\sum_{n=1}^{\infty}\sum_{p=1}^{\infty}
\frac{\phi_n(\xi)\phi_p(\xi)\tilde{\Phi}_{np}\Re(A_p)}{M_n^2M_p 
|A_n|^2 |A_p|^2} + {\cal O}(\tilde{\bm{\Phi}}^2) \right]. 
\end{multline}
In this expression, $\xi=x/L$, 
$\tilde{\Phi}_{np}$ is given by \eqref{Lambda0mat}
and the quantity $A_n=\Lambda_0(\omega)_{nn}/M_n$ reads
\begin{equation} \label{An}
A_n = \omega^2_n-\omega^2 (1+\lambda_n \Gamma(\omega))   
\end{equation}  
where $\omega_n=\sqrt{k_n/M_n}$ is the frequency in vacuum, 
$\lambda_n=M_{\!f}/M_n$ and $\Re(A_n)$ is the real part of $A_n$.
An expression similar to \eqref{specdensy2} can be derived for the 
slope of the deflection
%\eqref{specdensdy} 
provided $\phi_j(\xi)$ is replaced by $\phi'_j(\xi)$ and the overall prefactor
is divided by $L^2$.
%-----------------------------------------------------------------
\begin{figure}
\begin{center}
\scalebox{0.7}{\input{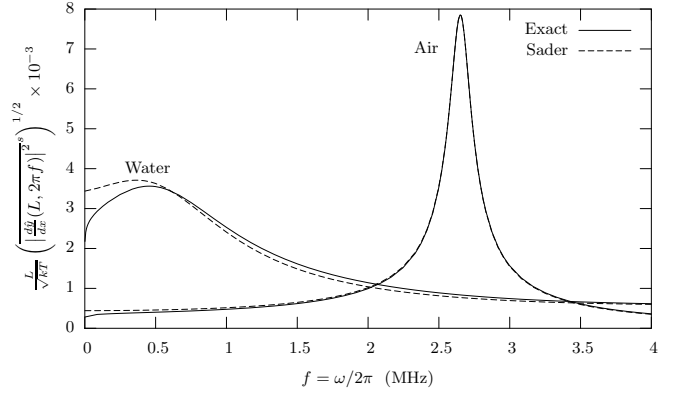}}
\caption{Spectral density of the deflection slope for a {\em bare} 
silicon nanobeam in air and water (solid lines) compared to 
Sader's expression \cite{theory_Sader} (dashed lines).}
\label{comp2sader}
\end{center}
\end{figure}
%------------------------------------------------------------------
% When $\bm{\Phi}=0$, the mass of modes $y_n$ 
% (see Eq. \eqref{Sadersmodes}) are all equal
% to the mass of the beam, $M_n=M$, and the first term of \eqref{specdensy2} 
% represents the {\em exact} 
% spectral density of the deflection of the bare beam. The 
% second term is of course zero.
% When $\bm{\Phi} \neq 0$, each mode $y_n$ acquires a different effective 
% mass,  
% $M_n=M+\Phi_{nn}$, and interacts with the others. In this case, the  
% first term of \eqref{specdensy2} essentially accounts for the mass 
% renormalization of the modes while the second takes into account their 
% coupling. As we will see below, mass renormalization is expected to be the 
% main contribution to the frequency shift of the bare beam peaks.  

Expression \eqref{specdensy2} is valid for any mass
distribution $\mu_l(x)$ along the beam. For a uniform layer with 
linear mass $\mu_l$, $\Phi_{np}=L \mu_l \delta_{np}$, and then
$\tilde{\Phi}_{np}=0$. All modes have the same effective mass, $M_n=M+L\mu_l$,
and are decoupled. Reinstating in \eqref{specdensy2}, the second term 
vanishes and we obtain the {\em exact} 
spectral density of a composite beam consisting of the original 
beam plus the layer. For molecules trapped on 
the beam at positions $\xi_i$, the mass profile given in 
Eq.\eqref{linmass1} leads to 
$\Phi_{np}=m\sum_i\phi_n(\xi_i)\phi_p(\xi_i)$ and Eq.\eqref{specdensy2} is valid 
up to first order in $m$ provided the frequency satisfies 
$M_{\!f} \Gamma_i(\omega) \gg 4 M_l$. 
Interestingly, if we 
assume $N$ molecules to be randomly scattered along the beam
in a uniform way and average $\Phi_{np}$ accordingly, we find
$\langle\langle \Phi_{np} \rangle\rangle = 
m\sum_i\int \phi_n(\xi_i)\phi_p(\xi_i)d\xi_i = Nm\delta_{np}$. As the 
total mass of the trapped molecules is small compared to the mass of the beam, in the 
first approximation, the average spectral density is the same as 
the spectral density of their average mass distribution---i.e. the 
spectral density of a uniform layer of mass $Nm$.   
     
In Fig. \ref{comp2sader}, we compare the bare beam
($\bm{\Phi}=0$) spectral density of the deflection slope at the tip of 
a rectangular silicon nanocantilever ($E=160$ Gpa, 
$\rho=2.33\times 10^3$ kg/${\rm m}^3$) to Sader's result
\cite{theory_Sader} in air and water. The beam dimensions
are $d\times b \times L=0.2\times 0.2\times 10 (\mu m)$. From  
\cite{Chon}, at $T=27^{o}$ C, the 
viscosities  are $\eta_{\rm air}=1.86\times 10^{-5}$, 
$\eta_{\rm water}=8.59\times 10^{-4}$ (kg/m/s) and the densities,  
$\rho_{\rm air}=1.18$, 
$\rho_{\rm water}=997$ (kg/${\rm m}^3$). Although different, 
Sader's formula can be shown to reduce to \eqref{specdensy2} provided 
$\lambda |\Gamma(\omega)| \ll 1$. 
This explains why the results are very
similar in air ($\lambda |\Gamma(\omega)|\simeq 0.06$ at resonance) 
while they start to differ in water 
($\lambda |\Gamma(\omega)|\simeq 25$ at resonance). 

%\section{Frequency shift}
%\subsection{Uniform layer}
%---------------------------------------------------------

{\it Frequency Shift:}
As stated earlier, when trapped molecules 
form a uniform layer, the {\em exact} spectral density of the beam deflection
is given by
\begin{equation} \label{specdensyUL}
\overline{\left|\hat{y}(x,\omega)\right|^2}^s =\frac{2k T M_{\!f}
\omega \Gamma_i(\omega)}{(M+M_l)^2} \, \sum_{n=1}^{\infty}\frac{
\phi^2_n(\xi)}{|A_n|^2},
\end{equation}
where $M_l$ is the mass of the layer and where $A_n$ is given in Eq.\eqref{An}
with $M_n=M+M_l$. When the peaks of Eq.\eqref{specdensyUL} are sharp enough, 
the hydrodynamic function is
almost constant in their vicinity and the resonant frequency satisfies 
the self-consistent equation
\begin{equation} \label{resfreqUL1}
\omega_{R,n}^2 = f(\omega_{R,n},M_n)\, k_n,\ \ 
f=\frac{1}{3}\frac{R+\sqrt{4R^2+3I^2}}{R^2+I^2}
\end{equation}   
where $R=M_n\!+\!M_{\!f}\Gamma_r$, $I=M_{\!f}\Gamma_i$ and 
$\Gamma_{r,i}\equiv \Gamma_{r,i}(\omega_{R,n})$. From the expression 
\eqref{resfreqUL1}, the mass and stiffness variations due to
the layer, $\delta M$ and $\delta k_n$, induce a relative frequency 
shift between a bare and a loaded beam:
\begin{equation} \label{freqshiftUL1}
\frac{\delta \omega_{R,n}}{\omega_{R,n}} = \frac{1}
{2-\omega_{R,n}\frac{\partial \ln f}{\partial \omega_{R,n}}}
\left[\frac{\delta k_n}{k_n}+\frac{\partial \ln f}{\partial M}
\delta M\right].
\end{equation} 
%-------------------------------------------------------------
\begin{figure}
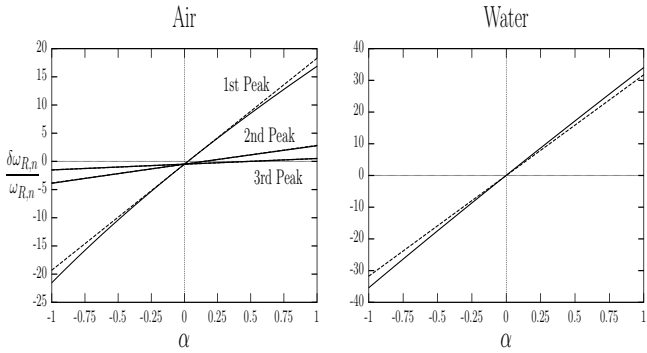

%\begin{center}
{\centering 
 \subfigure{\resizebox*{0.49\columnwidth}{0.49\columnwidth}{\input{Shift_Air.tex}}} 
 \subfigure{\resizebox*{0.49\columnwidth}{0.49\columnwidth}{\input{Shift_Water.tex}}}
\par}
%\scalebox{0.7}{\input{Shift_Air.tex}}
\caption{Relative frequency shift (in \%) in air and water vs the 
dimensionless stress, $\alpha=\sigma L^3/EI$ . 
$M_l=M/100$ and $E_l=0$. Other parameters are the same as in Fig. 
\ref{comp2sader}. Solid line: exact result from Eq. 
\eqref{specdensyUL}. Dotted line: linear approximation from
Eqs. \eqref{freqshiftUL1} and \eqref{deltakn}.}
\label{ShiftAir}
%\end{center}
\end{figure}
%--------------------------------------------------------------
Here, $\delta M = M_l$ and its prefactor in \eqref{freqshiftUL1} 
takes into account dissipative ($M_{\!f}\Gamma_i$) and 
fluid mass loading ($M_{\!f}\Gamma_r$) effects. 
According to \eqref{kn}, the two contributions to the 
stiffness $\delta k_n$ come from the bending rigidity, 
$E_b I_b \rightarrow E_b I_b \!+\! E_l I_l$, and from the surface
stress through the eigenvalue $\beta_n(\alpha)$. From \eqref{EI} and 
\eqref{betan}, we find 
\begin{equation} \label{deltakn}
\frac{\delta k_n}{k_n} = \frac{3 h E_l}{d E_b}+
\frac{2 T_{n} t_{n}+ 
\beta_{0,n}(T_{n}+t_{n}) }
{\beta_{0,n}^3(t_{n}-T_{n})}\alpha ,
\end{equation}
where $\beta_{0,n}$ is the $n$th root of $\cosh(\beta)\cos(\beta)+1=0$ and
where $T_{n}=\tanh\beta_{0,n}$, $t_{n}=\tan \beta_{0,n}$. The last term of
\eqref{deltakn} has been obtained from \eqref{betan} in perturbation. It is 
valid when $\alpha \ll n^2$ and vanishes
as $n\rightarrow \infty$. 

Using the same data as in Fig.~\ref{comp2sader}, 
we display in Fig.~\ref{ShiftAir} the
relative frequency shift in air (left) and water (right) versus the 
dimensionless surface stress $\alpha$. The exact shift is evaluated from
the spectral density \eqref{specdensyUL} and compared to its linear 
approximation \eqref{freqshiftUL1}. The layer mass has been arbitrarily fixed 
to 1\% of the beam mass and $E_l$ set to zero, hence the negative offset
observed in air at $\alpha=0$. For typical values of the 
surface stress, $\sigma \sim 10^{-2}{\rm J.m}^{-2}$ (see Wu {\it et al.} in ref. \cite{Majumdar01}),
$|\alpha| \lesssim 1$.
In air, $\lambda |\Gamma(\omega_{R,n})|\ll 1$, 
and $f \sim 1/M$. Then, $\partial \ln f/\partial M \sim -1/M$, 
$\partial \ln f/\partial \omega_{R,n} \sim 0$, and
we recover the usual frequency shift for a linear oscillator in vacuum. As seen
on the left panel, the first peak is the most sensitive to $\alpha$. The 
deviation of the data from the linear result \eqref{deltakn} 
indicates that the condition $\alpha \ll n^2$ with $n=1$ becomes violated.
This effect disappears for the second and third peaks that are less 
sensitive to $\alpha$. In water (right panel), a single broad peak 
occurs. Eq. \eqref{resfreqUL1} loses its accuracy but the 
frequency shift \eqref{freqshiftUL1} derived from it is 
still acceptable. The contribution of $\partial \ln f/\partial M$ becomes 
negligible while $\partial \ln f/\partial \omega_{R,n}\sim -
\partial \ln \Gamma_i/\partial \omega_{R,n}$ becomes
important, hence the increase in the slope of the relative frequency
shift versus $\alpha$ in water compared to air.  

In conclusion, we treat the stochastic response of biofunctionalized nanomechanical
cantilevers with a generalized fluctuation-dissipation relation. In a viscous fluid like water, the resonance frequency shift for a continuous distribution of biomolecules on the cantilever surface is dominated by surface stress rather than mass loading.

% 
%---------------------------------------------
%\subsection{Particles trapped at the tip}
%We end this section by evaluating the frequency shift generated 
%by a single or group of particles with mass $m$ 
%trapped at the tip of the cantilever ($\xi=1$). For simplicity, the deflection
%is assumed to be measured at the tip too. Then, using $\phi_n(1)=2(-1)^n$ 
%(valid for all cantilever modes), we see that $\Phi_{nl}=4m(-1)^{n+l}$ and 
%\begin{equation} \label{specdensy3}
%\overline{\left|\hat{y}(L,\omega)\right|^2}^s \!\!=\frac{8k T M_{\!f}
%\omega \Gamma_i(\omega)}{M_t^2} \sum_{n=1}^{\infty}\frac{\left[1   
%\!+\! \frac{32\omega^2m}{M_t}{\displaystyle \sum_{p\neq n}}
%\frac{\Re(A_p)}{
%|A_p|^2}\right]
%}{|A_n|^2}\!. 
%\end{equation} 
%--------------------------------


\begin{thebibliography}{1}
\bibitem{footnote}
As $\Gamma_i (\omega) \rightarrow 0$ 
when $\omega \rightarrow \infty$, $M_{\!f} \Gamma_i(\omega) \gg 4 M_l$
is violated at some critical frequency beyond which Eq. 
\eqref{chipert} does not hold. A ``non-perturbative'' treatment of the inversion of 
$\Lambda(\omega)$ is then in order. It can be done approximately by 
analytically inverting the $2\times 2$ or $3\times 3$ sub-block 
of $\Lambda(\omega)$ surrounding the small diagonal element.
\end{thebibliography}
\end{document}